\documentclass[prd,twocolumn,showpacs,reprint,preprintnumbers,nofootinbib]{revtex4-1}
\input{header}

\begin{document}

%============================================================
\title{Co-interacting dark matter and conformally coupled light scalars}
%============================================================

\author{Philippe Brax}
\email{Philippe.brax@ipht.fr}
\affiliation{Institut de Physique Th$\acute{\textrm{e}}$orique, Universit$\acute{\textrm{e}}$ Paris-Saclay, CEA, CNRS, F-91191 Gif-sur-Yvette Cedex, France}

\author{Carsten van de Bruck}
\email{c.vandebruck@sheffield.ac.uk}
\affiliation{Consortium for Fundamental Physics, School of Mathematics and  Statistics, University of Sheffield, Hounsfield Road, Sheffield, S3 7RH, UK}

\author{Sebastian~Trojanowski}
\email{strojanowski@camk.edu.pl}
\affiliation{Astrocent, Nicolaus Copernicus Astronomical Center Polish Academy of Sciences, ul.~Rektorska 4, 00-614, Warsaw, Poland}
\affiliation{National Centre for Nuclear Research, Pasteura 7, 02-093 Warsaw, Poland}

\begin{abstract}
The increasing observational pressure on the standard cosmological model motivates analyses going beyond the paradigm of the collision-less cold dark matter (DM). Since the only clear evidence for the existence of DM  is based on gravitational interactions, it seems particularly fitting  to study them  in this sector where extensions to the standard model can be naturally introduced. A promising avenue can be obtained using modifications of the space-time metric coupled to DM and induced by the presence of a new ultra-light scalar field $\phi$. The $\phi$ field can contribute to the DM density and can couple all the matter species universally, including additional heavy DM particles.  We present a simple  two-component DM model employing derivative conformal interactions between the two DM species. This can simultaneously: 1) guarantee the  necessary symmetries to stabilize the dark species, 2) predict a subdominant thermal relic density of the heavy DM component, and 3) alleviate small-scale structure tensions of the cold DM scenario due to the possible co-interactions in the dark sector. The scenario is highly predictive with future observational prospects ranging from the Large Hadron Collider (LHC) to gravitational-wave searches, and  can be generalized to more rich and realistic dark sector models.
\end{abstract}

\renewcommand{\baselinestretch}{0.85}\normalsize
\maketitle
\tableofcontents
\renewcommand{\baselinestretch}{1.0}\normalsize

\section{Introduction}

The existence of cold dark matter (CDM) with its far-reaching implications for the large-scale structure of the Universe has long been one of the crucial parts of the cosmological standard model ($\Lambda$CDM). This dominant paradigm has been further strengthened by particle physics, in which an ideal CDM candidate has been proposed in numerous scenarios predicting the existence of cosmologically-stable weakly interacting massive particles (WIMPs) (see e.g. Refs~\cite{Arcadi:2017kky,Roszkowski:2017nbc} for recent reviews). It has been shown that WIMPs could be thermally produced in a way that is insensitive to the details of the evolution of the Universe before their \textsl{freeze-out}. The electroweak-size dark matter (DM) couplings and masses naturally lead to the correct DM relic abundance, $\Omega_{\textrm{DM}}^{\textrm{tot}} h^2\simeq 0.12$~\cite{Aghanim:2018eyx}, without the need for additional fine-tuning of the parameters describing the very early evolution of the Universe. This remarkable and fairly model-independent observation is known as the \textsl{WIMP miracle}.

In recent years, however, the WIMP DM models and the $\Lambda$CDM scenario have been under growing pressure from astrophysical and cosmological observations, cf. Ref.~\cite{Perivolaropoulos:2021jda} for a recent review. In the DM sector of the Universe, the corresponding challenges can be divided into two main categories: \textsl{1) small-scale structure problems of CDM} and \textsl{2) the lack of signal of WIMP DM in dedicated searches}. In the former case, the current observations differ from the expectations from simulations~\cite{Kroupa:2010hf,Weinberg:2013aya}, although some tensions could at least partially be weakened by e.g. taking into account the impact of the baryonic feedback, cf. Refs~\cite{Somerville:2014ika,Salucci:2018hqu,Vogelsberger:2019ynw} for recent reviews. A particularly well-discussed issue is the so-called \textsl{core vs cusp problem} concerning the DM density profiles in central regions of galaxies~\cite{Moore:1994yx,Flores:1994gz,Walker:2011zu}. Further hints of beyond the CDM cosmology come from analyzing the distributions of satellite galaxies around the Milky Way~\cite{Klypin:1999uc,Moore:1999nt,Kroupa:2004pt,BoylanKolchin:2011de,BoylanKolchin:2011dk} and, more recently, from an excess of small-scale gravitational lenses observed in galaxy clusters~\cite{Meneghetti:2020yif}.

On the other hand, the lack of convincing signals in DM direct detection (DD) and indirect detection (ID) experiments also contradicts theoretical predictions in an increasing number of popular WIMP models. In particular, the simplest  scenarios predicting thermally-produced $\chi$ with the mass $m_\chi\lesssim 100~\gev$ have been ruled out by the Fermi-LAT $\gamma$-ray searches coming from dwarf spheroidal galaxies (dSphs)~\cite{Fermi-LAT:2016uux}, while even stronger constraints can be derived based on the observation of the Galactic Centre (GC)~\cite{Abazajian:2020tww}. Further bounds can be obtained from the observations of the antiproton and positron fluxes reaching the AMS-02 detector (see e.g.~\cite{Aguilar:2016kjl,Cuoco:2016eej,John:2021ugy}). 

These observations have triggered a growing interest in alternative DM scenarios that could explain better the data. The prime examples of such theories are fuzzy or ultra-light bosonic DM $\phi$ with a mass $m_\phi\sim 10^{-22}~\ev$~\cite{Hu:2000ke,Hui:2016ltb,Lee:2017qve,Ferreira:2020fam} and the self-interacting DM (SIDM) models with $\sigma/m\sim 1~\cm^2/\g$~\cite{Spergel:1999mh}. Among them, the vanilla fuzzy DM model might already be ruled out by the observations of the Lyman-$\alpha$ forest~\cite{Irsic:2017yje,Armengaud:2017nkf,Kobayashi:2017jcf,Nori:2018pka}. Notably, though, this could be avoided for  scalar field masses $m_\phi\gtrsim 10^{-21}~\ev$ in models with repulsive self-interacting ultra-light DM~\cite{Fan:2016rda}. On the other hand, the DM self-interactions are constrained by observations of galaxy clusters and cluster mergers, see Ref.~\cite{Tulin:2017ara} for review and Ref.~\cite{Andrade:2020lqq} for a recent  constraint from cluster strong lensing, $\sigma/m< 0.065~\cm^2/\g$. In the SIDM case, the fit to astrophysical data requires then the relevant cross section to be made velocity-dependent, so that this scenario could avoid the aforementioned bounds and address the issues arising at smaller scales~\cite{Kaplinghat:2015aga}. However, it remains challenging in such simple  models to  fit simultaneously  the small-scale structure data and predict the correct DM relic density, as well as to avoid current bounds from cosmology and direct searches~\cite{Hambye:2019tjt}. 

In this study, we argue that similar solutions to the small-scale structure issues can also be obtained in a co-interacting DM regime~\cite{Liu:2019bqw} with a two-component DM: a subdominant contribution from a heavy WIMP-like $\chi$ and the dominant one associated with an ultra-light scalar field $\phi$. To this end, we employ a derivative coupling between both dark species, which can e.g. emerge from a minimal conformal modification to  the Einstein metric~\cite{Kugo:1999mf,Kaloper:2003yf} that preserves Lorentz invariance and causality~\cite{Bekenstein:1992pj}. In a nutshell, we assume that matter couples to a Jordan metric which is conformally related to the Einstein metric, i.e. the one for which the Einstein-Hilbert term of General Relativity is normalised by  Newton's constant. It can therefore naturally affect all the matter species, preferably in a universal way such that it conforms to the weak equivalence principle and constitutes a simple portal between the visible and dark sectors of the Universe. Importantly, these interactions preserve a  shift symmetry for the ultra-light $\phi$ and allow for introducing additional $\mathbb{Z}_2$ symmetries stabilizing both $\chi$ and $\phi$ on cosmological scales. Last but not least, the derivative conformal $\phi$ interactions can naturally lead to the correct thermal abundance of the subdominant heavy DM component $\chi$~\cite{Trojanowski:2020xza}.\footnote{See also Refs~\cite{Brax:2020gqg,Brax:2021gpe} for recent discussion about the conformal and disformal couplings leading to efficient DM production in the \textsl{freeze-in} mechanism.}
\medskip

This paper is organized as follows. In \cref{sec:model} we introduce the derivative conformal couplings between the SM and the dark species and describe the relevant constraints. We discuss the relic abundance of both DM components in \cref{sec:relicdensity}. In \cref{sec:solitioncointDM}, we illustrate how the co-interacing DM regime can be realized in the model under study. Other phenomenological implications of the presence of such conformal couplings are presented in \cref{sec:pheno}. We conclude in \cref{sec:conclusions}. \Cref{app:Boltzmann} provides a more detailed discussion of the Boltzmann equation for the ultra-light species. In \cref{app:solitons}, we analyze loop-induced self-interactions that appear in the scenario under study and the resulting soliton size.

\section{Derivative conformal couplings of light scalars and WIMP DM\label{sec:model}}

The effective interaction terms  between the ultra-light scalar field $\phi$ and heavy dark fermion $\chi$, as well as with the SM, are given by
\begin{equation}
\mathcal{L}= \frac{(\partial\phi)^2}{M_{\textrm{SM/DM}}^4}\,T_{\textrm{SM/DM}},
\label{eq:confL}
\end{equation}
where $(\partial\phi)^2= \partial_\mu\phi\,\partial^\mu\phi$ and the energy-momentum tensor for $\chi$ reads
\begin{equation}
T_{\textrm{DM}}^{\mu\nu} = -\frac{i}{2}\left[\bar{\chi}\gamma^{(\mu}\partial^{\nu)}\chi - \partial^{(\mu}\bar{\chi}\gamma^{\nu)}\chi\right].
\label{eq:Tmunu}
\end{equation}
One finds the relevant expressions for the Standard Model (SM) by using respective covariant derivatives. We have defined the trace of the energy-momentum tensor as $T_{\textrm{DM}}=(T_{\textrm{DM}})^{\mu}_\mu$. In the following, we will assume that ultra-light scalars $\phi$ couple universally to all matter species, $M_{\textrm{DM}}=M_{\textrm{SM}}\equiv M$. Notably, by coupling pairs of $\phi$ and $\chi$ species, the interaction in \cref{eq:confL} also allows for further discrete symmetries to be imposed that could stabilize both dark species on cosmological time scales.

Another important property of the interaction Lagrangian, \cref{eq:confL}, is that it preserves the shift symmetry for $\phi$, which helps in stabilizing the scalar mass with respect to quantum corrections from its couplings to the much heavier field $\chi$ and to the SM. Therefore, this symmetry can only  be very softly broken, e.g., by introducing a small but finite mass for the scalar, $m_\phi$, or by other terms in the respective scalar potential $V(\phi)$. Below, we assume that all such shift-symmetry-breaking terms are much suppressed and they do not affect the DM phenomenology under study. The only notable exception in this context is that a possible small shift of $\phi$ from the minimum of the potential induced by these terms can lead to late-time oscillations of the scalar field that behaves as a non-relativistic matter, similar to the axion field, for review see Ref.~\cite{Baer:2014eja} and references therein. 

As mentioned above, the interaction Lagrangian that we have introduced can naturally arise if the $\phi$ field is assumed to drive a slight modification of the space-time metric. 
Indeed, let us assume that the matter action reads
\begin{equation}
    S_m\equiv S_m(\psi_i, \tilde g_{\mu\nu}),
\end{equation}
where the matter fields are generically denoted by $\psi_i$. These fields include the dark matter field $\chi$ as one of the matter fields interacting with the scalar $\phi$. The metric $\tilde g_{\mu\nu}$ is the Jordan metric and differs from the Einstein metric $g_{\mu\nu}$ which appears in the Einstein-Hilbert term of the gravitational action. A general decomposition of the Jordan metric was given by Bekenstein \cite{Bekenstein:1992pj} and reads
\begin{equation}
    \tilde g_{\mu\nu}= C(\phi,X) g_{\mu\nu} + D(\phi,X) \partial_\mu \phi \partial_\nu \phi,
\end{equation}
where $X = -(1/2)\,g^{\mu\nu}\partial_\mu\phi\,\partial_\nu\phi$.
Such simple conformal and disformal transformations between the metric in the Einstein and the Jordan frames preserve Lorentz invariance and causality. In this paper we assume that the disformal interaction vanishes, i.e. $D(\phi,X)=0$ and therefore focus on\footnote{Due to a too large suppression of the non-relativistic $\phi-\chi$ scattering cross section~\cite{Trojanowski:2020xza}, a universal disformal coupling to all the matter species does not reproduce the co-interacting DM regime for the models under study. Similar conclusions are  true for the conformal interaction \cref{eq:confL} if the heavy DM component $\chi$ is a complex scalar field.}   
\begin{equation}
\tilde{g}_{\mu\nu} = C(\phi,X)\,g_{\mu\nu}\ .
\label{eq:metricconf}
\end{equation}
Such transformations affect geodesics but also generate $\phi$ couplings to the matter fields. Indeed, defining the matter energy-momentum tensor as 
\begin{equation}
    T^{\mu\nu}=-\frac{2}{\sqrt{-\tilde g}} \frac{\delta S_m}{\delta \tilde g_{\mu\nu}},
\end{equation} 
the interaction terms appear in the series expansion of the action around the vacuum, which corresponds to $\phi=0$ and the Minkowski metric $g_{\mu\nu} = \eta_{\mu\nu}$~\cite{Kugo:1999mf,Kaloper:2003yf}, i.e.
\be 
\delta S_m= -\frac{1}{2} \int d^4x \delta \tilde g_{\mu\nu} T^{\mu\nu}
\ee
where $\tilde g_{\mu\nu}= \eta_{\mu\nu} + \delta \tilde g_{\mu\nu}$.
The particular operator in  \cref{eq:confL} can be obtained from the function $C(\phi,X)\equiv C(X) \approx 1+4X/M^4$~\cite{Brax:2016kin} corresponding to $\delta \tilde g_{\mu\nu}= -2 [(\partial \phi)^2/M^4]\,\eta_{\mu\nu}$ which applies to all the SM fields and DM. See also discussion in \cite{Aviles:2010ui}.

On the other hand, the interaction in \cref{eq:confL} can also be considered independent of this motivation. Notably, \cref{eq:confL} corresponds to  dimension-8 operators that leads to  cross sections strongly growing with the interaction energy. One expects that above a certain  energy scale of order $M$, the model should be replaced by a more fundamental theory. In the following, however, we will focus on the interactions energies and values of $M$, for which the effective field theory (EFT) description in \cref{eq:confL} remains valid, cf. Ref.~\cite{Trojanowski:2020xza} and references therein for further discussion. 

The possible exception to this in our analysis is the treatment of the searches for $\phi$ missing-energy signatures at the Large Hadron Collider (LHC). Current such bounds are at the level of $M\gtrsim 200~\gev$~\cite{Aaboud:2019yqu}, cf. also Ref.~\cite{Brax:2016did}, which is below the typical center-of-mass collision energy at the LHC. For this reason, these constraints are not fully independent of the unknown UV completion of the simplified model. The impact of this completion is typically parameterized with an effective coupling $g_\ast$ that dictates the characteristic scale at which the EFT approach ceases to be valid, $\sqrt{\hat{s}}<g_\ast\,M$, where $\sqrt{\hat{s}}$ corresponds to the hard interaction. In particular, for $g_\ast\sim 1$, the LHC energies are too large to set bounds on the model within the EFT scenario. These bounds should then be treated in a model-dependent way that goes beyond the simplified approach and could become weaker. In the following, however, we will conservatively adopt the aforementioned value of the minimal constraint on $M$ that corresponds to $g_\ast> \pi^2$, cf. Ref.~\cite{Aaboud:2019yqu}, while we will indicate in our plots that the LHC bounds should be considered as approximate; see also Ref.~\cite{Cohen:2021gdw} for recent discussion on unitarity bounds in EFT. Importantly, independent lower bounds on $M$ could also be set based on the measurements of the $Z$ boson decays at the Large Electron Positron (LEP) collider. In particular, in Ref.~\cite{Brax:2015hma} such a discussion for a similar disformal derivative coupling and the $Z\to \mu^+\mu^-\phi\phi$ decay width has been given, which leads to independent bounds, $M \gtrsim \textrm{tens of }\gev$.

In a previous study~\cite{Trojanowski:2020xza}, we have identified a narrow region of the parameter space of the model, in which the aforementioned constraints can be satisfied and the heavy WIMP DM relic density can be partially or even entirely driven by the $\chi$ conformal coupling to $\phi$. This corresponds to the following approximate ranges of the $\chi$ mass and the universal conformal mass $M$
\begin{align}
\label{eq:mchirange}
100~\gev \lesssim &\ m_\chi\ \lesssim 1~\tev,\\
200~\gev \lesssim &\ M\ \lesssim 1.2~\tev,\\
 & m_\chi \lesssim \ M\ .
\label{eq:Mrange}
\end{align}
In the current study, we extend this analysis by discussing signatures of the mixed $\phi\,+\,\chi$ DM scenario that go beyond the $\Lambda$CDM paradigm. Still, however, the above allowed ranges of the model parameters apply to our discussion, while they will become even more tightly constrained once we require $\chi$ to contribute subdominantly to the total DM relic density to reproduce the non-CDM behavior in the co-interacting DM regime. 

The presence of the dominant ultra-light bosonic DM component can, however, lead to additional bounds on the mass and couplings of the scalars. We note, though, that derivative interactions in \cref{eq:confL} are generally screened from the fifth force searches in the non-relativistic regime~\cite{Joyce:2014kja}. In addition, as we will see below, typical values of the scalar mass of  interest are of order $m_\phi\sim (10^{-14}-10^{-13})~\ev$ and lie much above the constraints from the Lyman-$\alpha$ forest~\cite{Irsic:2017yje,Armengaud:2017nkf,Kobayashi:2017jcf,Nori:2018pka}. In particular, these masses would  lead to solitonic structures with sizes much smaller than kpc. It is important to mention, however, that even heavier scalars can lead to observable effects by inducing superradiant instabilities around  spinning black holes (BHs)~\cite{Cardoso:2005vk,Dolan:2007mj}. These can lead to detectable gravitational waves emitted by bosonic condensates~\cite{Brito:2017wnc} or can manifest themselves as unexpected features in the BH spin-mass plane~\cite{Brito:2017zvb}. Currently such bounds exclude light scalar masses in the range between $\sim 10^{-13}$ and $10^{-11}~\ev$~\cite{Cardoso:2018tly,Ng:2020ruv}, as well as for lower masses, in the limited ranges around $m_\phi\sim 10^{-17}$ and $10^{-15}~\ev$~\cite{Wen:2021yhz}. These constraints remain outside of the region in the parameter space of the model relevant to our study.

While we focus only on the effective $\phi$ portal between $\chi$ and the SM, other interactions of heavy fermionic $\chi$s with the SM particles might emerge in more complete theoretical frameworks. We stress that such interactions would not affect our discussion as long as they are weaker than thermal in the early Universe. Notably, such conditions can easily be satisfied in many WIMP models and might, in fact, be preferred given increasingly more severe bounds from direct and indirect searches for DM. In addition, in order to simplify our discussion, we assume that $\chi$ self-interactions are suppressed, as it is the case for a typical collisionless CDM. The last condition is not strict and models with non-negligible $\chi$ interactions induced by other types of couplings could lead to additional signatures, on top of the ones discussed below. We leave such analysis for the future.

\section{The relic abundance of light and heavy DM components\label{sec:relicdensity}}

In this section, we discuss how the relic density of both DM components is obtained. We begin with briefly recapitulating the results of Ref.~\cite{Trojanowski:2020xza} with regards to the subdominant relic abundance of heavy $\chi$. We then discuss late-time oscillations of the dominant ultra-light DM component $\phi$, cf. Ref.~\cite{Brax:2019fzb}. 

\subsection{Subdominant heavy $\chi$ relic density}

The presence of the derivative conformal coupling between the dark species, cf. \cref{eq:confL}, allows the heavier particles $\chi$ to thermalize in the early Universe. This is mediated by light $\phi$s that are produced in the thermal plasma from their interactions with the SM. As discussed above, it is appealing to assume that $\phi$ couples universally to all the matter species (to the SM and $\chi$) via the same conformal coupling and the mass scale $M$. This settles both the $\chi$ and $\phi$ freeze-out temperatures that are approximately given by $T_{\chi,\textrm{fo}}\sim m_\chi/20$ and $T_{\phi,\textrm{fo}}\sim\gev$, respectively. 

For the parameter values of  interest, both freeze-out processes occur typically at temperatures below the one relevant for the electroweak phase transition, $T_{\textrm{EW}}$, but before the QCD phase transition characterized by $T_{\textrm{QCD}}$. The latter condition also allows one to avoid otherwise stringent bounds from the Big Bang Nucleosynthesis (BBN) from the $\phi$ contribution to the number of relativistic degrees of freedom. We then obtain $T_{\textrm{EW}}>T_{\chi,\textrm{fo}}>T_{\phi,\textrm{fo}}>T_{\textrm{QCD}}$.

In the non-relativistic limit, the derivative coupling  leads to the following $p$-wave suppressed annihilation cross section for the process $\chi\bar{\chi}\to\phi\phi$~\cite{Trojanowski:2020xza}
\begin{equation}
\sigma v \stackrel{v\ll 1}{\simeq} \frac{m_\chi^6}{32\pi\,M^8}\,v^2.
\label{eq:sigmaannpwave}
\end{equation}
The $\chi$ relic abundance is set by the assisted freeze-out mechanism~\cite{Belanger:2011ww}, and to a good approximation it is given by
\begin{equation}
\Omega_\chi h^2\sim (0.1)\,\sqrt{\frac{100}{g_{\ast}(x_{\chi,\textrm{fo}})}}\,
\left(\frac{x_{\chi,\textrm{fo}}}{20}\right)\,\left(\frac{10^{-9}\,\textrm{GeV}^{-2}}{\langle\sigma v\rangle}\right),
\label{eq:chiOh2}
\end{equation}
where $x_{\chi,\textrm{fo}}=m_\chi/T_{\chi,\textrm{fo}}$ and $g_{\ast}(x_{\chi,\textrm{fo}})$ is the number of relativistic degrees of freedom at $T_{\chi,\textrm{fo}}$. In particular, this can lead to a subdominant contribution from $\chi$ to the total DM relic density
\begin{equation}
\Omega_\chi h^2 = f\,\Omega_{\textrm{DM}}^{\textrm{tot}},\hspace{0.1cm}f\simeq 0.1\times \left(\frac{M}{300~\gev}\right)^8\left(\frac{300~\gev}{m_\chi}\right)^6.
\label{eq:chifracDM}
\end{equation}
In the following, we will rely on more precise calculations based on numerical solutions of the relevant set of Boltzmann equations.

It is useful to note that the period of thermal equilibrium will also result in a leftover abundance of $\phi$ acting as hot DM (HDM). This contribution to the $\phi$ relic density is, however, strongly suppressed for the aforementioned range of the dark scalar mass
\begin{equation}
(\Omega_\phi h^2)_{\textrm{HDM}} \sim 10^{-16}\,\left(\frac{100}{g_{\ast s}(x_{\phi,\textrm{fo}})}\right)\,\left(\frac{m_\phi}{10^{-13}\,\textrm{eV}}\right)\ll\Omega_{\textrm{DM}}^{\textrm{tot}},
\label{eq:Oh2relativistic}
\end{equation}
where $x_\phi = m_\phi/T_{\phi,\textrm{fo}}$. Therefore, it does not affect the formation of the large-scale structure of the Universe and it plays a negligible role in our analysis. 

\subsection{The dominant relic abundance of ultra-light $\phi$\label{sec:oscillations}}

The dominant relic abundance of $\phi$ is settled at a later epoch, when the rapid oscillations of the $\phi$ field begin and after both dark species decouple from the SM. Such oscillations are initially effectively frozen, as they are damped by the large value of the Hubble rate, $H\gg m_\phi$. In the radiation dominated (RD) epoch (i.e. for $T\gtrsim 1~\ev$), this rate decreases with the temperature as $H= (\pi/\sqrt{45})\,T^2/M_{\textrm{Pl}}$. Hence, for the scalar masses of order  $m_\phi\sim 10^{-14}~\ev$, the temperature at which we expect that $H\sim m_\phi$ and the background $\phi$ field begins to oscillate is of order $T_{\textrm{osc}} \sim \mathcal{O}(10~\mev)$. Since then, the harmonic oscillations of the $\phi$ field resemble non-relativistic cold DM, although they provide a subdominant contribution to the total energy-momentum budget of the Universe until the epoch of matter-radiation equality.

In the following, we will not deal any  longer with the model building aspects of the scalar field dynamics. In particular, the construction of such model would require us to design the interaction potential $V(\phi)$ and to explain the origin of the misalignment which would eventually lead to the scalar oscillations reproducing the phenomenology of CDM. In the following we will simply assume that the scalar potential admits a minimum where the scalar field has a mass $m_\phi$ and that the amplitude of the oscillations around this minimum are such that this constitutes a dominant part of the DM energy content. 

\section{Co-interacting DM\label{sec:solitioncointDM}}

Having established the two-component DM scenario with the dominant contribution from ultra-light $\phi$s, we will now discuss the  impact of the interactions between the two dark species on DM haloes. We focus on the scatterings between the heavy $\chi$ particles and $\phi$ species existing in the form of a non-relativistic gas.\footnote{Instead,  $2\to 3$ and $3\to 2$ processes involving larger number of $\phi$s can naturally be forbidden in the model under study by imposing discrete symmetries consistent with the interaction Lagrangian \cref{eq:confL}. We then neglect their impact in the following.} These could induce SIDM-like heat transfer toward the galactic cores. The relevant cross section in the non-relativistic regime ($E_\phi\sim m_\phi$) is given by~\cite{Trojanowski:2020xza}
\begin{widetext}
\begin{equation}
\label{eq:scatconf}
\sigma_{\phi\chi} \simeq \frac{m_\chi^5\,m_\phi}{2\pi\,M^8},\\
\simeq (2\times 10^{-31}\,\gev^{-2})\times \left(\frac{m_\phi}{3\times 10^{-14}~\textrm{eV}}\right)\left(\frac{m_\chi}{300~\gev}\right)^5\left(\frac{300~\gev}{M}\right)^8\ .
\end{equation}
\end{widetext}
As can be seen, the predicted value of $\sigma_{\phi\chi}$ is very small. This is expected for the derivative conformal coupling for which the cross section grows rapidly only in the relativistic limit, $E_\phi\gg m_\chi$.\footnote{If the conformal coupling in \cref{eq:confL} arises from modified gravity, additional $\phi$ self-interaction terms $(\partial\phi)^4$ and $(\partial\phi)^2\phi^2$ should be induced from the energy-momentum tensor of the $\phi$ field~\cite{BeltranJimenez:2018tfy}. Such scattering rates will, however, be much suppressed with respect to the $\phi\chi$ co-interactions due to a much lower center-of-mass energy in the $\phi\phi$ interactions. In fact, the strength of these $\phi$ self-interactions will be of similar order to the loop-induced term discussed in \cref{app:solitons}.}

However, even tiny co-interaction cross sections characteristic for the late-time stage of the evolution of the Universe can lead to observable effects. This primarily relies on a huge enhancement factor in the scattering rates from a large occupation number of the light bosonic species. After averaging over the fast oscillation of the scalar field, the magnitude of this enhancement is driven by both the scalar mass and its typical velocity, $\langle\mathcal{N}_\phi\rangle \simeq (\rho_\phi/m_\phi)\,\lambda_\phi^3$, where de Broglie wavelength of the scalar particles reads $\lambda_\phi = 2\pi/(m_\phi v_0)$. We note here that even though both DM species are produced non-relativistically and their momenta become further suppressed due to the expansion of the Universe, they are eventually accelerated to galactic speeds during the structure formation process. For the typical values of the galactic DM density $\rho_{\textrm{DM}}\sim 0.1\,M_{\odot}/\textrm{pc}^3$ and velocity $v_0\sim 10^{-3}\,c$~\cite{Walker:2009zp,Oh:2010ea}, and for the light scalar mass for $m_\phi = 3\times 10^{-14}~\textrm{eV}$, one obtains the value of the corresponding phase-space density function of order $\mathcal{N}^\phi\sim \textrm{a few}\times 10^{58}$. Notably, the Bose enhancement becomes weaker for a growing $\phi$ velocity $v_0$, as expected due to a diminishing de Broglie wavelength of the field.

\begin{figure}[t]
    \centering
\includegraphics[width=0.47\textwidth]{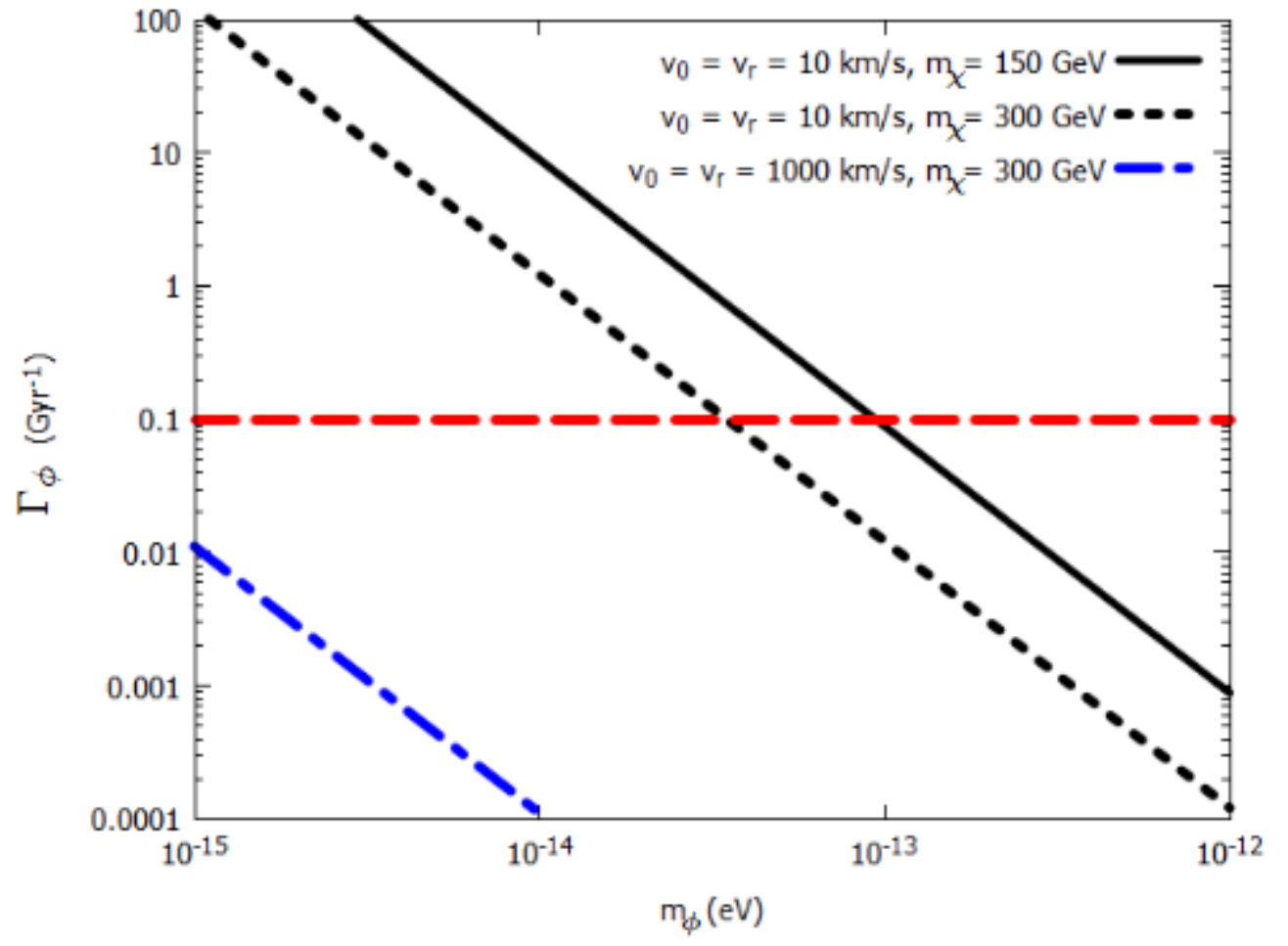}  
\caption{The effective interaction rate $\Gamma_\phi$ of the light scalars $\phi$ due to their scatterings with the heavy DM component $\chi$ as a function of the scalar mass $m_\phi$. We show with the black solid (dotted) line the value of $\Gamma_\phi$ for $m_\chi = 150~\gev$ ($300~\gev$) and the DM velocities characteristic for dwarf galaxies $v_r \simeq v_0 \simeq 10~\km/\textrm{s}$. The blue dash-dotted line corresponds to $m_\chi = 300~\gev$ and larger velocities $v_r\simeq v_0\simeq 1000~\km/\textrm{s}$ relevant for clusters of galaxies. All the lines are shown for $\rho_{\textrm{DM}}^{\textrm{tot}} = 0.1 M_\odot/\textrm{pc}^3$. The horizontal red dashed line indicates the value of $\Gamma_\phi = 0.1~\textrm{Gyr}^{-1}$.}
\label{fig:gammaphi}
\end{figure}

A more detailed estimate of the interaction rate can be obtained by solving the corresponding Boltzmann equation, which also takes into account the effect of the forward-backward suppression in the collisional kernel for the $\phi\chi$ scatterings~\cite{Liu:2019bqw}. This is due to a tiny momentum change of the heavy $\chi$ particles in each individual collision with $\phi$, which results in  roughly the same probability for $k_1\to k_2$ and inverse $k_2\to k_1$ transitions. The relevant suppression factor reads $\mathcal{N}^\chi_{k_1} - \mathcal{N}^\chi_{k_2} \sim \mathcal{N}^\chi_k\,(m_\phi/m_\chi)$. The effective co-interaction rate for the dominant $\phi$ DM component is then given by (see \cref{app:Boltzmann} for further discussion)
\begin{equation}
\Gamma_\phi \simeq n_\chi\,\langle\sigma_{\phi\chi} v_r\rangle\,\langle\mathcal{N}^\phi\rangle\,\left(\frac{m_\phi}{m_\chi}\right),
\label{eq:Gammaphi}
\end{equation}
where $v_r$ is the relative velocity between the two dark species and $\langle \sigma v \rangle$ describes the scattering cross section averaged over the DM velocity distribution. By substituting \cref{eq:chifracDM,eq:scatconf} into \cref{eq:Gammaphi} we obtain
\begin{widetext}
\begin{equation}
\Gamma_\phi\sim (0.1~\textrm{Gyr}^{-1})\,\left(\frac{3\times 10^{-14}~\ev}{m_\phi}\right)^2\,\left(\frac{300~\gev}{m_\chi}\right)^3\,\left(\frac{\rho^{\textrm{tot}}_{\textrm{DM}}}{0.1 M_\odot/\textrm{pc}^3}\right)^2\,\left(\frac{v_r}{10~\textrm{km}/\textrm{s}}\right)\,\left(\frac{10~\textrm{km}/\textrm{s}}{v_0}\right)^3,
\label{eq:Gammaphiconf}    
\end{equation}
\end{widetext}
where we have used the vanilla DM interaction rate of $\Gamma_\phi\sim 0.1~\textrm{Gyr}^{-1}$ required by the SIDM solutions to the core vs cusp problem in dwarf galaxies~\cite{Tulin:2017ara}. We stress that the predicted subdominant $\chi$ DM relic density described by the factor $f_\chi$ in \cref{eq:chifracDM} is already implicitly taken into account in \cref{eq:Gammaphiconf}. We also assume that a relative velocity between the dark species is driven by the velocity dispersion, $v_r\sim v_0\sim 10~\textrm{km}/\textrm{s}$. As can be seen, the interaction cross section \cref{eq:scatconf} reproduces the correct value of the total $\phi$ interaction rate $\Gamma_\phi$ for $m_\phi = (\textrm{a few})\times 10^{-14}~\ev$ and $m_\chi$ of order several hundred $\gev$. In \cref{fig:gammaphi}, we show the $\phi$ interaction rate as a function of $m_\phi$ obtained for $m_\chi = 150$ and $300~\gev$, as well as for DM velocities of order $10$ and $10^3~\km/\textrm{s}$. The latter velocities are more characteristic of galaxy clusters and lead to much suppressed co-interaction rates.

The interaction rate in \cref{eq:Gammaphiconf} does not explicitly depend on the conformal mass scale $M$. This is because a decrease in $M$ leads to a simultaneous increase in the cross section $\sigma_{\phi\chi}$ and a drop in the fraction of the heavy DM component $f_\chi$. Both effects compensate each other in $\Gamma_\phi\sim f_\chi\,\sigma_{\phi\chi}$, as can be deduced from \cref{eq:chifracDM,eq:scatconf}. It should be noted, however, that an implicit impact of $M$ on the results is still expected. The larger the conformal mass becomes, the more $\chi$ DM-dominated scenario one obtains, $f_\chi\to 1$. Eventually, the $f_\chi$ fraction goes beyond the $\phi$-dominance regime, in which the interaction rate can be defined with \cref{eq:Gammaphi}. We discuss below that this results in more stringent bounds from cluster mergers or leads to a standard CDM-like behavior of DM haloes. For this reason, in the following, we will require $f_\chi\lesssim 0.1$. Given the EFT validity and LHC bounds discussed in \cref{sec:model}, such $f_\chi$ bound limits the available $\chi$ masses to a relatively narrow range of
\begin{equation}
170~\gev\lesssim m_\chi\lesssim 300~\gev\ .
\end{equation}

In fact, for most of the available parameter space of the model, we obtain $0.1 \gtrsim f_\chi\gtrsim 0.01$, i.e. the heavy DM component corresponds to $(1-10)\%$ contribution to the total DM density. We stress again a remarkable fact that, already in the minimal effective model, the obtained value of $f_\chi$ in \cref{eq:chifracDM} can naturally predict astrophysically relevant value of $\Gamma_\phi$ in \cref{eq:Gammaphiconf}. This also results in specific predictions for the mass of the light scalar $\phi$, which can be related to $m_\chi$ via \cref{eq:Gammaphiconf} by requiring $\Gamma_\phi\sim 0.1~\textrm{Gyr}^{-1}$ in typical dwarf galaxies. By decreasing the masses of individual DM components, one obtains larger values of $\Gamma_\phi$, which is expected due to the growing number densities of both dark species. The approximate range of the dark scalar masses for which both $f_\chi$ and $\Gamma_\phi\sim 0.1~\textrm{Gyr}^{-1}$ can be fitted corresponds to $3\times 10^{-14}~\ev\lesssim m_\phi < 7\times 10^{-14}~\ev$. 

Last but not least, we stress the dependence of $\Gamma_\phi$ in \cref{eq:Gammaphi} on DM relative velocity $v_r$ and the velocity dispersion $v_0$. As discussed above, this allows one to obtain substantial $\phi$ interaction rates in DM-dominated dwarf or low-surface brightness (LSB) galaxies, in which $v_0\sim v_r$ is, typically, of order tens of $\textrm{km}/\textrm{s}$. Notably, such galaxies remain the prime laboratory to test DM properties, given the corresponding lower expected baryonic feedback on DM haloes, see also Ref.~\cite{DiPaolo:2019eib} for recent analysis of possible non-standard properties of DM haloes of LSB galaxies. On the other hand, in galaxy clusters with characteristic velocities $v_0\sim \mathcal{O}(1000~\textrm{km}/\textrm{s})$, the interaction rate in \cref{eq:Gammaphi} is much suppressed. As a result, the model satisfies bounds from cluster strong lensing~\cite{Andrade:2020lqq} and ellipticity~\cite{Miralda-Escude:2000tvu}. Similar such bounds from the galactic halo shape of NGC 720 with $v_0\sim 340~\textrm{km}/\textrm{s}$ can also be evaded~\cite{Feng:2009hw}; see also Refs~\cite{Peter:2012jh,Rocha:2012jg,Tulin:2017ara} and references therein for a more detailed and updated discussion.

Further important constraints on dark matter self-interactions come from observations of cluster mergers; see Ref~\cite{Molnar2016review} for review. Here, the $\phi$ interaction rate is also suppressed due to large typical DM velocities. Additional suppression might arise in this case for DM species with the relative velocity between the two colliding DM haloes that exceeds the typical velocity dispersion in each of them. In this case, the collision kernel in the Boltzmann equation for $\phi-\chi$ scatterings is no longer enhanced by the final-state Bose enhancement factor for $\phi$ since the typical velocity of light scalars after the collision is of order $v_r\gg v_0$. The relevant distribution then gains a Maxwell-Boltzmann suppression factor $\sim \exp(-v_r^2/v_0^2)$, where the effective temperature of the dark $\phi$ species is driven by $kT_\phi\sim m_\phi v_0^2$. The suppression is of order $10^{-7}$ for the velocities characteristic for the Bullet Cluster, cf. Ref.~\cite{Liu:2019bqw}. Similar discussion holds for DM substructures within the main galactic~\cite{Spergel:1999mh} or cluster~\cite{Gnedin:2000ea} haloes and the respective bounds can also be avoided in the co-interacting regime.

\begin{figure*}
    \centering
    \includegraphics[width=0.49\textwidth]{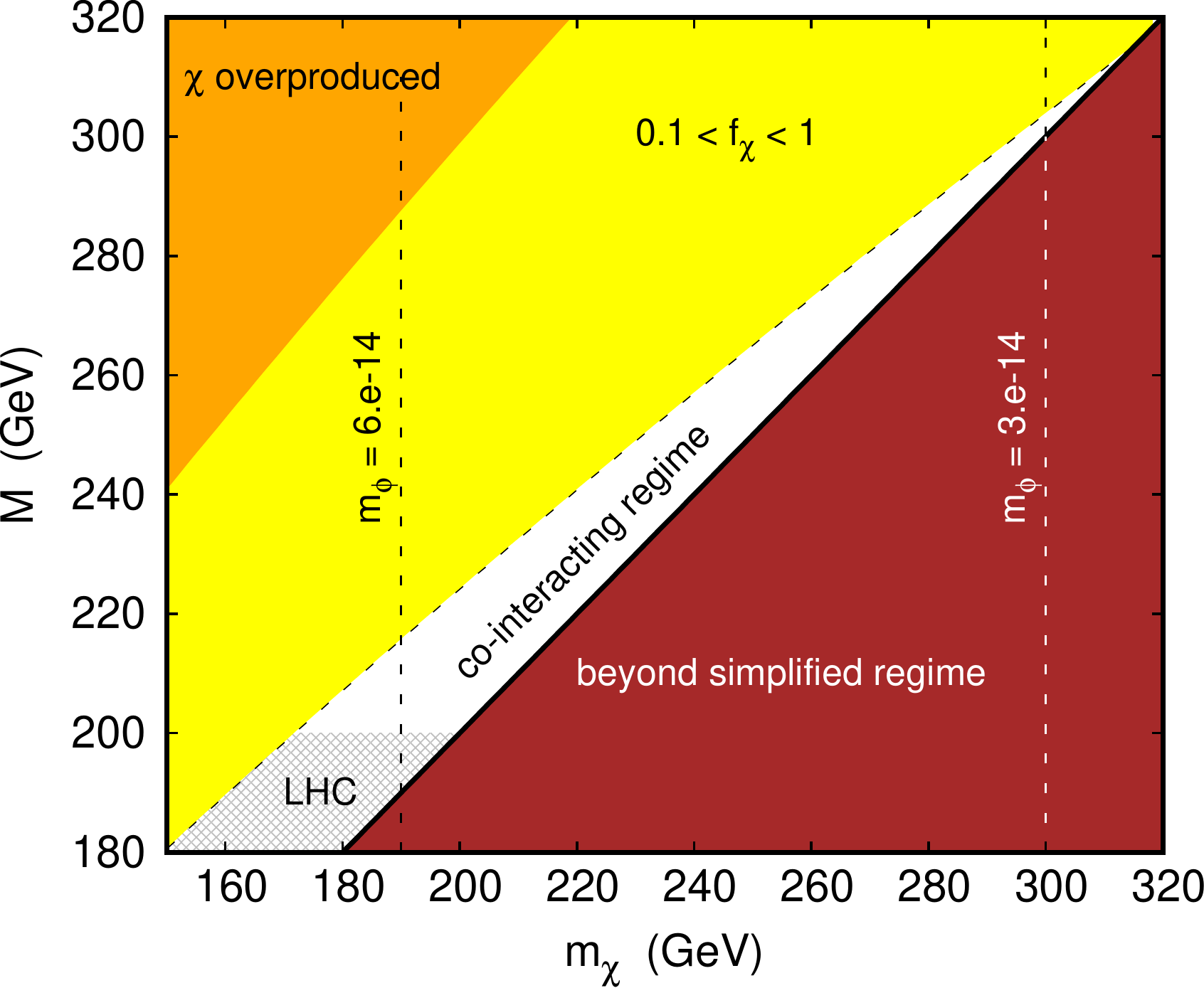}
\hfill
    \includegraphics[width=0.49\textwidth]{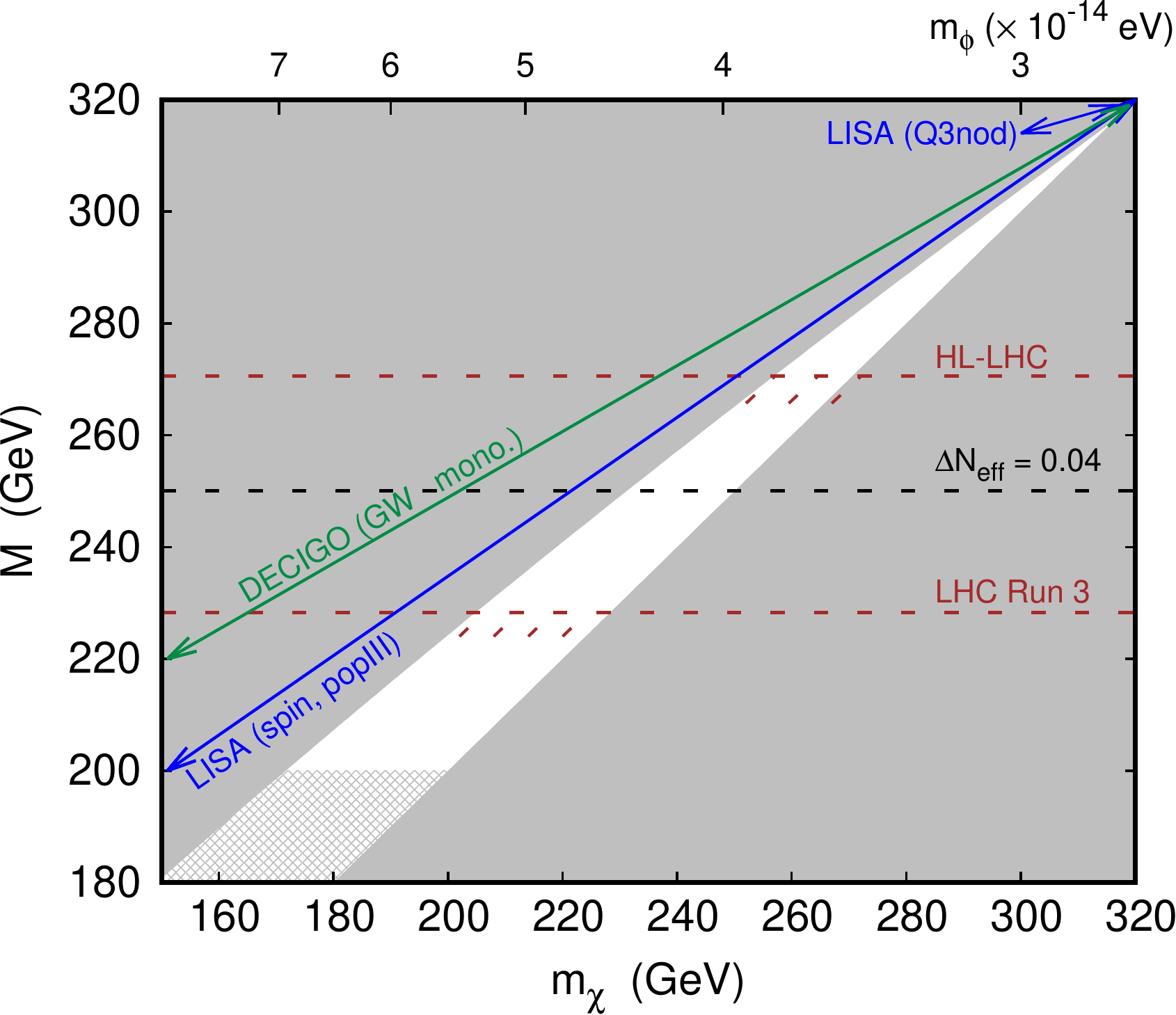}
\caption{\textsl{Left}: The preferred region in the parameter space of the simplified model presented in the $(m_\chi,M)$ plane, in which the co-interacting DM regime is found with $\Gamma_\phi\sim 0.1~\textrm{Gyr}^{-1}$ in dwarf galaxies. The relevant values of the light scalar mass $m_\phi$ are also indicated with vertical lines. The colorful regions correspond to the approximate EFT, LHC bounds, as indicated in the plot. In the yellow-shaded region, the heavy DM component $\chi$ has a sizeable abundance that exceeds $10\%$ of the total DM density and the standard cold DM regime is reproduced (see text). In the orange region, the thermal relic density of the $\chi$ DM component overcloses the Universe. \textsl{Right}: Similar to the left panel, but focusing on the future observational prospects. The co-interacting DM regime is obtained in the white region in the plot. Expected future bounds on the maximum value of the conformal mass $M$ are shown with the dashed brown lines. The black dashed line corresponds to the effective number of additional, $\phi$-induced, relativistic degrees of freedom in the early Universe equal to $\Delta N_{\textrm{eff}} = 0.04$. The colorful diagonal arrows describe the expected sensitivity of the future gravitational-wave detectors to ultralight scalars with the mass $m_\phi$, which is related to $m_\chi$ and $M$ by \cref{eq:Gammaphiconf}. We show the scale relevant for $m_\phi$ on top of the plot. These bounds are not sensitive to the precise value of the conformal mass $M$ (see text for details).}
\label{fig:cointeracting}
\end{figure*}

So far we have focused on the effective interaction rate of light $\phi$ species mediated by their interactions with $\chi$. These interactions will, however, also affect the distribution of heavy DM particles. These interact more often than the $\phi$ species. By the time one ultra-light scalar $\phi$ interacts with a single heavy $\chi$ particle, each such $\chi$ particle interacts with as many as $n_\phi/n_\chi \sim (m_\chi/m_\phi)\,(1/f_\chi)$ different target $\phi$s, where the difference in the interaction rates is driven by the ratio of the number densities of both species. Importantly, however, a single $\phi\chi$ scattering process will typically have only a minor impact on the $\chi$s momentum, $\delta p_\chi \sim m_\phi v \ll m_\chi v$, and multiple such scatterings are required for a substantial effect to occur. In an individual halo, we expect that only after roughly $N_{\textrm{col.}} \sim (p_\chi/\delta p_\chi)^2~\sim (m_\chi/m_\phi)^2$ collisions the combined momentum change of $\chi$ will be of order $p_\chi$ due to a random walk in the momentum space. In other words, for astrophysically-relevant effect, one needs to calculate the rate of $N_{\textrm{col.}}$ consecutive interactions of $\chi$, which introduces a suppression factor of $(m_\phi/m_\chi)^{-2}$ with respect to the total $\chi$ interaction rate. Taking into account both effects, i.e. the enhancement by $n_\phi/n_\chi$ and the suppression by $1/N_{\textrm{col.}}$, one expects $\Gamma_\chi^{\textrm{dwarf}} \sim (m_\phi/m_\chi)\,\Gamma_\phi^{\textrm{dwarf}}$ and $\chi$s remain effectively collisionless in dwarf galaxies. As a result, the scenario, in which the heavy DM component dominates, is then characterized with CDM-like behavior with basically no impact of the co-interacting regime on astrophysical observables. 

Focusing further on the $\chi$-dominance scenario, we note that the relevant interaction rate $\Gamma_\chi$ could be much increased for smaller values of $m_\phi\ll 10^{-14}~\ev$. In this case, however, the subdominant fraction of $\phi$ would be characterized by a very large value of $\Gamma_\phi$ in dwarfs and it would no longer reproduce the vanilla SIDM framework. In addition, further bounds in this case are expected from cluster mergers. Here, the $\phi$ and $\chi$ species can come from different haloes with $v_r\gg v_0$ moving in opposite directions. The aforementioned random walk suppression factor should then be replaced with a factor linear in $(m_\phi/m_\chi)$. This is because, in this case, consecutive interactions have directional preference and more quickly add up to the total momentum exchange of the order of $p_\chi$. The actual value of the suppression factor depends on how central is the cluster collision. As a result, one obtains $ \Gamma_\chi^{\textrm{cl.mer.}}\gg \Gamma_\chi^{\textrm{dwarf}}$, and the co-interaction rate of the heavy DM component would explicitly violate the bounds on DM interactions from observations of cluster collisions. We conclude that the $\chi$-dominance scenario with a non-negligible $\phi$ abundance cannot predict non-CDM-like properties of DM haloes and simultaneously be reconciled with the astrophysical data. Below, we will then focus on the $\phi$-dominance case with $f_\chi\lesssim 0.1$.

We summarize our results in the left panel of \cref{fig:cointeracting}. This has been obtained by numerically solving the Boltzmann equations relevant for the production of the subdominant $\chi$ DM in the early Universe and by evaluating the collision term in the Boltzmann equation corresponding to $\phi\chi$ co-interactions at later epoch. In the plot, we show in the $(m_\chi,M)$ plane the available region in the parameter space of the simplified model, in which the co-interacting DM regime leads to $\Gamma_\phi\sim 0.1~\textrm{Gyr}^{-1}$ in dwarf galaxies. We also indicate there the relevant values of the light $\phi$ mass. The approximate EFT and LHC bounds are also shown in the plot which follows the discussion in \cref{sec:model}. In the orange region, we predict from \cref{eq:chiOh2} too large a DM relic density of $\chi$ that would overclose the Universe.

The yellow-shaded region in the plot corresponds to scenarios with a larger contribution from the heavy DM component, $0.1 \lesssim f_\chi\lesssim 1$. We stress again that for fixed $m_\phi$ and $m_\chi$, increasing the conformal mass $M$ does not affect the $\phi$ interaction rate, $\Gamma_\phi\sim f_\chi\sigma_{\phi\chi}\sim \textrm{const}$. Therefore, the larger the value of $M$ is along vertical lines in \cref{fig:cointeracting}, the more CDM-like scenario one obtains due to increasing the $\chi$ fraction of the total DM abundance. This behavior of the model in the yellow-shaded region could be partially changed by mildly decreasing values of $m_\phi$ to $(10^{-15}-10^{-14})~\ev$ and, therefore, by increasing $\Gamma_\phi$ in \cref{eq:Gammaphiconf}. Still, however, in this case, only a fraction of DM that corresponds to $\phi$ species would behave similarly to SIDM, while the heavy component $\chi$ would be CDM-like. Notably, in this case, further bounds on light $\phi$s can be deduced from the superradiant instabilities around massive BHs, see \cref{sec:model}. Specifically, for $m_\phi\sim 10^{-15}~\ev$, this could be in contradiction with the fitted mass and spin value of the black hole in the tidal disruption event (TDE) 3XMM J215022.4-055108~\cite{Wen:2021yhz}, while for $m_\phi\sim 10^{-16}~\ev$ further bounds arise from the fitting the supermassive black hole mass and spin in the active galactic nuclei (AGN) in the nearby galaxy NGC 4051~\cite{Denney:2009kw,Patrick:2012ua}. Importantly, as discussed above, lowering the ultra-light scalar mass would also lead to too large $\chi$ co-interaction rate that violates the constraints from cluster mergers, unless the $\phi$ abundance was much suppressed. We conclude that, while the yellow-shaded region in the left panel of \cref{fig:cointeracting} is not excluded, there is relatively little room there to reproduce the non-CDM behavior of the co-interacting regime of our interest.

Whilst in our simplified scenario we assume no effective self-interactions between the $\chi$ particles, we briefly comment on the interesting phenomenology of less-simplified scenarios, in which such sizeable $\chi\chi$ interaction rates could be present. In particular, for $f_\chi\lesssim 0.1$, even large self-interaction cross section $\sigma_{\chi\chi}\gtrsim 10~\textrm{cm}^2/\textrm{g}$ can avoid bounds from cluster mergers, cf., e.g., Ref.~\cite{Randall:2008ppe} for the relevant Bullet Cluster constraints. The impact of such interactions on the $\chi$ distribution could, however, have interesting consequences for galactic subhaloes traveling through the main DM halo. In this case, efficient $\chi$ self-interactions could result in a stripping of the subhalo from the heavy DM component~\cite{Spergel:1999mh}, such that effectively $f_\chi^{\textrm{subh.}}\ll 1$, especially in less massive subhaloes with $M_{\textrm{subh.}}\lesssim 10^9 M_\odot$~\cite{Vogelsberger:2012ku}. This, in turn, also suppresses the co-interaction rate of the dominant $\phi$ component, since $\Gamma_\phi\sim f_\chi^{\textrm{subh.}}$. The subhalo would then become more CDM-like. 

In connection to this, we note that it has been reported~\cite{Zavala:2019sjk} that the vanilla SIDM models are in tension with the observed diversity in DM density profiles in ultra-faint Milky Way (MW) satellite galaxies. This is due to predicted too low subhalo densities in vanilla SIDM that are hard to reconcile with the data. Instead, this tension could be alleviated in the CDM scenario, in which only tidal disruption in the MW disk is taken into account without additional DM self-interactions. The expected subhalo densities could then become larger, as suggested by observations.\footnote{See also Refs~\cite{Zolotov:2012xd,GarrisonKimmel2019} for further discussions in relation to subhalo disruptions and the \textsl{too-big-to-fail} problem.} As discussed above, in the model of  interest with sizeable $\chi\phi$ co-interaction rates, if additional $\chi\chi$ self-interactions were present, the small subhaloes could be effectively stripped from the subdominant $\chi$ self-interacting DM. This would have minor effect on the total subhalo density, while it would become even more $\phi$-dominated and CDM-like. This effect could then allow for better explanation of the aforementioned diversity in ultra-faint galaxies, whilst maintaining the successful predictions of SIDM-like scenarios in more massive dwarfs. The impact on the subhalo densities could be further enhanced by possible gravothermal collapse~\cite{Lynden-Bell:1968eqn} of low-velocity DM haloes~\cite{Zavala:2019sjk,Turner:2020vlf}. A detailed analysis of this effect would require dedicated $N$-body simulations that could resolve the interplay between both DM components.

Finally, we note that, in the model under study, self interactions of ultra-light $\phi$s would also be generated at the loop level, e.g., with the exchange of heavy DM species $\chi$. However, in the parameter region of  interest, the resulting repulsive self interactions of $\phi$ are highly suppressed. Hence, they could support the existence of $\phi$ solitonic structures in dense regions of galaxies with characteristic size not larger than order $1~\cm$ that will have no astrophysical relevance. We discuss this in more detail in \cref{app:solitons}.

\section{Other phenomenological implications\label{sec:pheno}}

Besides having an impact on the current astrophysical observations, the two-component DM model that we have presented  will also lead to distinct signatures in future searches. We present several such prospects in the right panel of \cref{fig:cointeracting} and discuss them below. 

The search for conformally-coupled light scalars at the LHC could strongly constrain the available parameter space of the model during the upcoming data-taking periods. In the plot, we present the approximate future  bounds on the conformal mass $M$ that correspond to the search for $t\bar{t}+\slashed{E}$ signature with $150~\ifb$ and $3~\iab$ of integrated luminosity characteristic of the LHC Run 3 and High-Luminosity LHC (HL-LHC) era, respectively. This search provides the most promising bounds on $M$ for light scalars coupled via \cref{eq:confL}, since $\phi$s couple most strongly to the heaviest SM fermions. The constraints have been obtained assuming the value of the effective coupling equal to $g_\ast\simeq \pi^2$, cf. discussion in \cref{sec:model}. As can be seen, values of $M$ up to $\sim 270~\gev$ could be probed in the coming years in this collider search. We note, however, that these constraints could become significantly weaker for decreasing $g_\ast$, which reflects their dependence on the UV completion of the simplified scenario.

Independent future probes of ultra-light scalars will be possible in searches for gravitational-wave signatures induced by superradiant instabilities around massive BHs. In  Fig. \ref{fig:cointeracting} , we first represent such bounds based on the predicted ``holes'' in the BH spin-mass plane~\cite{Brito:2017zvb} that could be seen in the future Laser Interferometer Space Antenna (LISA) data~\cite{LISA:2017pwj}. These correspond to blue diagonal two-headed arrows in \cref{fig:cointeracting} that will constrain $m_\phi$ independent of $m_\chi$ and $M$. In order to present these bounds in the $(m_\chi,M)$ plane, we then assume that both the dark sector masses are related by the requirement of fitting the co-interaction rate, $\Gamma_\phi\sim 0.1~\textrm{Gyr}^{-1}$ in \cref{eq:Gammaphiconf}. We show the relevant range of $m_\phi$ in the upper horizontal axis in the plot. Since the gravitational wave (GW) signal induced by light scalars with $m_\phi\sim (10^{-14}-10^{-13})~\ev$ is expected to occur at the higher end of the planned frequency band of LISA, the detection prospects depend on the assumed BH population model~\cite{Klein:2015hvg}. The best reach could be obtained for the ``light-seed'' model denoted as popIII, in which the massive BHs are expected to grow from relatively light high-redshift seeds with masses of order a few hundred $M_\odot$. Instead, if the initial seed population was shifted toward larger masses (the Q3nod model, as well as the popular Q3 model not shown in the plot), the detection prospects in LISA would remain significantly worse.

An alternative $\phi$ detection strategy is based on direct emission of nearly monochromatic GWs by the bosonic condensate around the BH~\cite{Brito:2017wnc}. For the masses of $m_\phi$ of interest, however, the relevant signal will typically remain beyond the reach of both Advanced LIGO~\cite{LIGOScientific:2014pky} and LISA, as it would correspond to the frequency band gap between them, $f\sim (0.1-10)~\textrm{Hz}$. A possible exception, in this case, could be the most nearby sources ($z\sim 0.001$) that could be seen in Advanced LIGO, while further improvement in detection prospects is expected for the future Einstein Telescope~\cite{Maggiore:2019uih}. Interestingly, though, the relevant frequency band could also be very well covered by the proposed space-based gravitational wave antenna DECIGO~\cite{Kawamura:2011zz} and its pathfinder B-DECIGO~\cite{Nakamura:2016hna}. In this case, even very distant individual sources with $z\sim 3$ could lead to observable signals. In addition, the GW emission from  bosonic condensates could lead to a detectable stochastic gravitational-wave background. We indicate this in \cref{fig:cointeracting} with the two-headed green diagonal arrow. It should be noted, however, that the exclusion bounds derived based on superradiant instabilities rely on the assumption that the conditions around the massive BHs allow for the formation and growth of the bosonic condensate.

The late-time decoupling of light scalars $\phi$ in the early Universe could also result in their contribution to an effective number of relativistic degrees of freedom $\Delta N_{\textrm{eff}}$. It could then affect BBN or Cosmic Microwave Background (CMB) predictions by changing the expansion rate of the Universe. For the light scalars, however, this effect is typically mild of order $1\sigma$ observation in future CMB Stage 4 surveys~\cite{Abazajian:2019eic}, cf. discussion in \cite{Trojanowski:2020xza}. This is expected for spin-$0$ species decoupling before the QCD phase transition. In the right panel of \cref{fig:cointeracting}, we show  the predicted value  $\Delta N_{\textrm{eff}}\simeq (4/7)\,(10.75/g_\ast(T_{\textrm{dec}}))^{4/3}\simeq 0.04$ for $M\simeq 250~\gev$, which is also characteristic for the entire range of $M$ shown in the plot. This is due to a moderate dependence of the decoupling temperature of $\phi$ on the conformal mass, $T_{\phi,\textrm{dec}}\sim M^{8/7}$, and only small changes in $g_\ast(T)$ in the relevant range of this temperature $T_{\phi,\textrm{dec}}\gtrsim \gev$, cf. Ref.~\cite{Trojanowski:2020xza} for extended discussion.

The universal coupling of the light scalar to all the matter species will also induce $\phi$ interactions with the SM neutrinos. These could play an important role in the early Universe due to their possible impact on primordial density fluctuations affecting CMB observables and large-scale structure (LSS) predictions~\cite{Wilkinson:2014ksa,Olivares-DelCampo:2017feq}. The relevant bounds on the present value of the scattering cross section can be as low as $\sigma_{\phi\nu}\lesssim 10^{-56}~\textrm{cm}^2$ for the tiny $\phi$ DM mass, $m_\phi\sim 10^{-14}~\textrm{eV}$~\cite{Wilkinson:2014ksa,Mosbech:2020ahp}. This is assuming that the cross section is constant with the temperature, while the bound becomes even more stringent, $\sigma \lesssim 10^{-68}~\textrm{cm}^2$,  when the cross section grows quadratically with the temperature, $\sigma \sim T^2$. The conformal coupling considered in this paper leads to a cross section which can be deduced from the expression relevant for the $\phi$ scattering of the fermionic DM species $\chi$~\cite{Trojanowski:2020xza}
\begin{equation}
\sigma_{\phi\nu}\simeq \frac{1}{2\pi}\,\frac{E_\nu\,(m_\nu^2+2\,E_\nu\,m_\phi)^2\,m_\phi}{M^8}\ .
\label{eq:phinu}
\end{equation}
Assuming that $E_\nu\sim T\lesssim 10~\mev$, at which temperature the $\phi$ field starts to oscillate, we observe $m_\nu^2+2\,E_\nu\, m_\phi\sim m_\nu^2\sim (0.1~\textrm{eV})^2$. In this regime, $\sigma_{\phi\nu}\simeq (1/2\pi)\,E_\nu\,m_\nu^4\,m_\phi/M^8$ and the cross section depends linearly on $E_\nu\sim T$. As expected, in the non-relativistic regime, $E_\nu\simeq m_\nu$, \cref{eq:phinu} resembles \cref{eq:scatconf} obtained for the $\phi$ scattering of the heavy DM $\chi$ species. We find that the present value of the $\phi-\nu$ scattering cross section is orders of magnitude below the current bounds, $\sigma \sim 10^{-121}~\textrm{cm}^2$ for $m_\nu\sim 0.1~\textrm{eV}$. However, we should stress that the corresponding interaction rate in the Boltzmann equation can be affected by a very large Bose enhancement factor for the $\phi$ species, $\mathcal{N}_\phi\sim \textrm{a few}\times 10^{58}\times (10^{-3}/v_0)^3$. In fact, this could bring it right in the ballpark of cosmological relevance, in between the aforementioned bounds for the constant and $T^2$-dependent scattering cross sections, where it could also affect other cosmological observables, cf. discussions in \cite{Wilkinson:2014ksa,Hooper:2021rjc}. We leave a careful analysis of this effect for  future studies dedicated to cosmological simulations. 

The above phenomenological aspects of the model rely on various possible signatures related to the existence of ultra-light scalars $\phi$. In a simplified model, the heavy DM component $\chi$ is secluded and couples to the SM only via the $\phi$ portal, which remains suppressed in the non-relativistic regime. Therefore, it is beyond the reach of current and future direct detection searches in underground detectors. As far as indirect searches are concerned, we stress that the present-day annihilation rates of $\chi\chi\to\phi\phi$ are $p$-wave suppressed. In addition, the produced secondary flux of boosted $\phi$s remains undetectable on Earth with the characteristic scattering cross section off protons of order $\sigma_{\phi p}\sim (0.01~\textrm{ab})\times (m_\chi/1~\tev)^3$~\cite{Trojanowski:2020xza}, i.e., typically much below attobarn, where we have used $E_\phi = m_\chi$ after the annihilation process. Therefore, the simplified model under study also avoids these bounds. 

Last but not least, while we focus on the simplified scenario, the presence of additional sub-weak couplings between $\chi$ and the SM in more complete models would not typically affect our conclusions. These could, however, lead to separate phenomenological signatures of $\chi$ in both direct and indirect searches that would only very mildly be affected by the presence of the conformally coupled light scalars. We note, however, that the relevant signal rates would be suppressed for the subdominant $\chi$ DM component with $f_\chi\sim (0.01-0.1)$.

\section{Conclusions\label{sec:conclusions}}

The increasingly growing pressure from the lack of vanilla WIMP DM signal and possible deviations from the $\Lambda$CDM cosmological model has recently led  to a growing interest in alternative scenarios. Such analyses, however, often become challenging when the origin of the DM relic density is simultaneously taken into account. In this case, to satisfy current bounds and to improve the fit to the available observational data, one often has to  modify early periods in the cosmological history of the Universe, which currently remain hidden from our searches, or one relies on non-thermal and somewhat fine-tuned DM production mechanisms.

In this study, we have discussed an alternative and generic way of addressing current problems in the cold DM scenario, which could be employed in many DM models predicting the existence of WIMP-like particles that avoid current observational direct and indirect bounds but struggle to satisfy the cosmological constraints on DM abundance. Interestingly, on top of the WIMP-like particles, this employs an ultra-light scalar field $\phi$, which is ubiquitous in cosmology. The interplay between the two dark species can alleviate tensions in current DM observations in a way similar to self-interacting DM models. Simultaneously, their more efficient interactions in the early Universe could play a crucial role in determining the relic density of the heavy field $\chi$.

Specifically, in our analysis, we have focused on the derivative conformal coupling between $\chi$ and $\phi$ fields and a similar  coupling between $\phi$ and the SM species. Notably, the presence of such interactions can naturally be motivated by minimal modifications introduced in the Einstein metric that preserve causality and Lorentz invariance. The conformal coupling of this type simultaneously allows for the shift symmetry, which secures the ultra-light scalar mass, and for discrete symmetries stabilizing both  dark species.

While a similar mechanism can determine the DM relic density and its observational properties in more general and realistic dark sector models, we used a simplified framework. To this end, we focused on only the most essential fields and couplings needed to illustrate the idea. This allowed us to reduce the number of new parameters in such a scenario to just three: the two masses of the DM components, $m_\phi$ and $m_\chi$, and the coupling strength, which is described by the conformal mass $M$. We assumed that the latter is universal to all matter species and, therefore, satisfies the weak equivalence principle. We showed that, while this scenario is tightly constrained by astrophysical and collider bounds, the remaining small region in the parameter space of this effective model can simultaneously predict the astrophysically relevant co-interaction rate between the two dark species. This could then ease tensions present in the predicted small-scale structure of the Universe in the vanilla CDM scenario.

The model that we studied is characterized by a large predictive power with respect to future observations. All the three aforementioned parameters are determined up to a factor of a few. In particular, one expects $m_{\phi}\sim \textrm{a few }\times 10^{-14}~\ev$, while $m_\chi, M\sim \textrm{a few hundred}~\gev$. This translates into specific predictions in the proposed searches that range from future collider bounds at the LHC to gravitational wave searches in LISA and DECIGO observatories. Hints of new physics in such scenarios could also be detected by future CMB surveys. Instead, traditional ways of searching for DM signals in direct and indirect detection experiments are much less promising. This could also explain the lack of such signals in the searches up to this day.

We expect this generic discussion to hold in the presence of additional interactions between $\chi$ and the SM species in models with a more rich dark sector. This is provided that, in the early Universe, the dominant heavy $\chi$ DM interactions with the SM are via the $\phi$ portal, while other couplings lead to a sub-thermal interaction strength. Instead, since the derivative conformal interactions become much suppressed in the non-relativistic limit, the present-day phenomenology of $\chi$ in (in)direct searches will naturally be driven by these other couplings. In this case, however, these couplings could easily lie below current observational constraints. Still, the signatures of ultralight scalars $\phi$ will remain detectable in collider and gravitational-wave experiments.

The apparent problems of the $\Lambda$CDM scenario indicated by current observations motivates exploring new directions in particle cosmology. Possible interactions between ultra-light and much heavier dark species deserve special attention in these efforts. The derivative conformal operators offer a particularly attractive and well-motivated framework for such studies that can also lead to a unique combination of phenomenological effects. Understanding the full implications of such interactions, which could drive the dark matter sector of the Universe, will require further theoretical studies and numerical simulations. These can also be supported by future observational hints related to the existence of such ultra-light fields.

\acknowledgements CvdB is supported (in part) by the Lancaster–Manchester–Sheffield Consortium for Fundamental Physics under STFC grant: ST/T001038/1. ST is supported by the grant ``AstroCeNT: Particle Astrophysics Science and Technology Centre'' carried out within the International Research Agendas programme of the Foundation for Polish Science financed by the European Union under the European Regional Development Fund. ST is supported in part by the Polish Ministry of Science and Higher Education through its scholarship for young and outstanding scientists (decision no 1190/E-78/STYP/14/2019). 

\appendix

\subsection{Boltzmann equation for $\phi$\label{app:Boltzmann}}

A crucial quantity in our discussion is the co-interaction rate $\Gamma_\phi$ of the ultra-light scalar field $\phi$ with the heavy DM component $\chi$, cf. \cref{eq:Gammaphi,eq:Gammaphiconf}. It is useful to briefly recapitulate the most essential aspects of the discussion leading to the above expressions.

The phase-space density of the scalar field $\mathcal{N}_\phi$ follows the Boltzmann equation, which, in a general form, can be written as
\begin{equation}
\hat{L}[\mathcal{N}_\phi] = \hat{C}[\mathcal{N}_\phi]\ .
\end{equation}
Here, the Liouville operator $\hat{L}$ describes the time evolution of the system due to diffusion and external forces acting on $\phi$s in the expanding Universe. Instead, $\hat{C}$ is the \textsl{collision term}, which is associated with the interactions of $\phi$. As discussed in the main text, the crucial contribution to $\hat{C}$ comes from $\phi\chi$ co-interactions.

The collision term describing the $\mathcal{N}_\phi$ evolution due to the  $\phi(k_1)+\chi(p_1)\leftrightarrow \phi(k_2)+\chi(k_2)$ process with bosonic $\phi$ and fermionic $\chi$ reads
\begin{widetext}
\begin{align}
\int{\hat{C}[\mathcal{N}_\phi]\,\frac{d^3k_1}{(2\pi)^3}} = & \int{d\Pi_{k_1}d\Pi_{k_2}d\Pi_{p_1}d\Pi_{p_2}}\,(2\pi)^4\,\delta^{4}(k_1+p_1-k_2-p_2)\,|\mathcal{M}|^2\,\times\nonumber\\
& \hspace{2cm}\times\left\{\mathcal{N}_{\phi,k_1}\,\mathcal{N}_{\chi,p_1}\,(1+\mathcal{N}_{\phi,k_2})\,(1-\mathcal{N}_{\chi,p_2}) - (1+\mathcal{N}_{\phi,k_1})\,(1-\mathcal{N}_{\chi,p_1})\,\mathcal{N}_{\phi,k_2}\,\mathcal{N}_{\chi,p_2}\right\}
\end{align}
\end{widetext}
where $d\Pi_i = d^3p_i/(16\pi^3\,E_i)$ and $\mathcal{M}$ is the invariant matrix element for the considered process. Outside the dense regions in the galaxies, where, during the galaxy evolution, $\phi$ could form a Bose-Einstein condensate, its properties should resemble the ones of  a non-relativistic boson gas with a high occupation number, $\mathcal{N}_\phi\gg 1$. One can then simplify the product of phase-space densities
\begin{widetext}
\begin{equation}
\int{\hat{C}[\mathcal{N}_\phi]\,\frac{d^3k_1}{(2\pi)^3}} \simeq \int{d\Pi_{k_1}d\Pi_{k_2}d\Pi_{p_1}d\Pi_{p_2}}\,(2\pi)^4\,\delta^{4}(k_1+p_1-k_2-p_2)\,|\mathcal{M}|^2\,\left\{\mathcal{N}_{\phi,k_1}\,\mathcal{N}_{\phi,k_1}\,(\mathcal{N}_{\chi,p_1}-\mathcal{N}_{\chi,p_2})\right\},
\label{eq:Nchidiff}
\end{equation}
\end{widetext}
where we have omitted terms proportional to the product $\mathcal{N}_{\chi,p_1}\,\mathcal{N}_{\chi,p_2}\ll 1$ for the non-relativistic DM component $\chi$. Here, $\mathcal{N}_\chi \sim [\exp(E_\chi/T)+1]^{-1} \simeq \exp(-m_\chi/T)\ll 1$ for $m_\chi \gg T$. At a given temperature, the differential rate of the $\chi$ phase-space density change as a function of the $\chi$ three-momentum $p_\chi$ is given by $d\mathcal{N}_\chi/dp_\chi\simeq -\mathcal{N}_\chi\times p_\chi/(m_\chi\,T)$, where we have used $E_\chi = \sqrt{p_\chi^2+m_\chi^2}$ and $p_\chi\ll m_\chi$. For the typical momentum of $\chi$ in the DM halo, we observe $p_\chi^2 \sim m_\chi T$, so $d\mathcal{N}_\chi/dp_\chi \sim -\mathcal{N}_\chi/p_\chi$. In each individual $\phi\chi$ interaction, the characteristic momentum change of the heavier DM component $\chi$ is very small, $\delta p_\chi\sim m_\phi v_0\ll m_\chi v_0 \sim p_\chi$, as discussed in \cref{sec:solitioncointDM}. As a result, we can approximate the difference between phase-space densities of $\chi$ in \cref{eq:Nchidiff} as $|\mathcal{N}_{\chi,p_1}-\mathcal{N}_{\chi,p_2}|\simeq |(d\mathcal{N}_\chi/dp_\chi)\,\delta p_\chi| \sim \mathcal{N}_\chi\times \delta p_\chi/p_\chi\sim \mathcal{N}_\chi\,(m_\phi/m_\chi)$, i.e., the co-interaction rate in \cref{eq:Nchidiff} is affected by the forward-backward suppression factor proportional to the mass ratio between the ultra-light and heavy dark species, $m_\phi/m_\chi$~\cite{Liu:2019bqw}. This corresponds to the fact that in the collision term, both the direct and inverse scattering process are almost equally probable leading to the observed cancellation between the two contributions to $\hat{C}[\mathcal{N}_\phi]$.

The resulting collision term for $\phi$ can then be rewritten as
\begin{widetext}
\begin{equation}
\int{\hat{C}[\mathcal{N}_\phi]\,\frac{d^3k_1}{(2\pi)^3}} \simeq \frac{m_\phi}{m_\chi}\,\int{d\Pi_{k_1}d\Pi_{k_2}d\Pi_{p_1}d\Pi_{p_2}}\,(2\pi)^4\,\delta^{4}(k_1+p_1-k_2-p_2)\,|\mathcal{M}|^2\,\left\{\mathcal{N}_{\phi,k_1}\,\mathcal{N}_{\phi,k_1}\,\mathcal{N}_{\chi,p_1}\right\}.
\label{eq:Boltzsimpl}
\end{equation}
\end{widetext}
On top of the aforementioned forward-backward suppression, the collision term in \cref{eq:Boltzsimpl} differs from the standard result obtained for heavy non-relativistic DM species by the presence of an additional enhancement factor characteristic for large occupation number of the bosonic $\phi$ field in the final state, $\mathcal{N}_{\phi,k_2}$, cf. also Ref.~\cite{Davidson:2014hfa} for a similar discussion regarding axion self-interactions. The number density evolution of $\phi$ is then described by $dn_\phi/dt\propto n_\phi\,n_\chi\langle\sigma v\rangle_{\phi\chi}\langle\mathcal{N}_\phi\rangle\,(m_\phi/m_\chi)$. The relevant interaction rate is described by \cref{eq:Gammaphi}.

Importantly, while we have  focused on the $\phi\chi$ co-interaction rate, a large wavelength of $\phi$ could cause that the ultra-light $\phi$ species to simultaneously feel the impact of many heavy $\chi$ particles. If the $\phi$ wave functions after the scattering on a large number such $\chi$ species, $N_\chi\gg 1$, are in phase, this could lead to an additional enhanced coherent interaction rate of $\phi$. Naively, the coherent scattering off the system composed of $N_\chi$ heavy species would correspond to a significant enhancement of the scattering cross section driven by the large target mass in \cref{eq:scatconf}, $m_\textrm{tar}\sim N_\chi m_\chi$. This would only partially be compensated for by the suppression in the number density of such heavy targets. However, we note that this simple estimate goes beyond the validity regime of \cref{eq:scatconf} for the cross section obtained in the EFT approach, where we have assumed $m_{\textrm{tar}}\lesssim M\sim m_\chi$. In fact, we expect the scattering cross section to be regularized at high center-of-mass collision energies in the UV complete scenario such that such a strong growth in the scattering rate will be absent in the model, and we do not treat it in the main discussion. 

\subsection{Creation and size of $\phi$ DM solitons\label{app:solitons}}

In the main text, we have focused on the dominant co-interactions between light and heavy dark species present in the model under study. We have also noted that the de Broglie wavelength of the $\phi$ field with $m_\phi\sim 10^{-14}~\textrm{eV}$ is much below the kpc scale characteristic for fuzzy DM with the mass of order $10^{-22}~\textrm{eV}$. However, in the presence of sizeable repulsive $\phi$ self-interactions, even much heavier fields can lead to large solitonic structures in DM dense regions in galaxies~\cite{Fan:2016rda,Brax:2019fzb}. While we have \textsl{a priori} not introduced such interactions in our scenario, they will be induced at the loop level by the exchange of heavy $\chi$ fields, cf. \cref{fig:loopdiagram}. Below, we estimate the size of $\phi$ solitonic structures induced this way.

\begin{figure}[t]
    \centering
\includegraphics[width=0.27\textwidth]{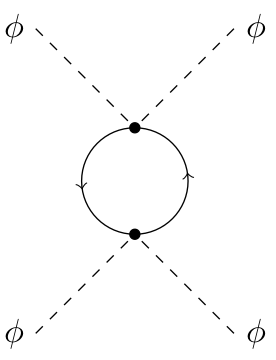}  
\caption{The Feynman diagram for the loop-induced self interactions of ultralight scalars $\phi$ with the exchange of the heavy DM component $\chi$.}
\label{fig:loopdiagram}
\end{figure}

Once we include the term that vanishes on shell, the trace of the energy-momentum tensor for the fermionic $\chi$ field is given by
\begin{equation}
(T_{\textrm{DM}})^{\mu}_{\mu} = 2m_\chi\,\bar{\chi}\chi - i\,\bar{\chi}\slashed{\partial}\chi.
\end{equation}
Substituting this into \cref{eq:confL}, we evaluate the relevant fermionic loop, which gets contributions that are up to quartically divergent. This corresponds to the fact that our effective model should be replaced with a more fundamental theory at the energy scale of order $M\gg m_\phi$. Although the effective model is \textsl{a priori} non-renormalizable, it could still be made predictive in the low-energy regime characteristic for non-relativistic $\phi$ self-interactions, cf. Ref.~\cite{Burgess:2007pt} for review. To this end, we isolate the log-divergent contributions by dimensional regularization and set higher order divergences to zero.\footnote{We have explicitly verified that similar order of magnitude estimates can be obtained by introducing an example auxiliary cutoff function to regularize the scattering amplitude, $f(\Lambda) = [\Lambda^2/(q^2-\Lambda^2)]^n$ with $n=4$, which generalizes the Pauli-Villars regularization obtained for $n=2$.} In the $\widebar{\textrm{MS}}$ scheme and in the non-relativistic limit with $p_\phi\sim m_\phi$, we obtain
\begin{equation}
\mathcal{L} \supset -\alpha(\Lambda)\,\frac{m_\chi^4}{M^8}\,(\partial\phi)^4 + \ldots,
\label{eq:confLapprox}
\end{equation}
where $\alpha(\Lambda) = (3/4\pi^2)\,\log{(\Lambda^2/m_\chi^2)}$ and $\Lambda\gtrsim m_\chi$. We have neglected the terms proportional to higher powers of the ultralight scalar mass $m_\phi$. The overall minus sign in front appears due to the presence of the fermionic loop. The resulting $\phi$ self-interactions are then repulsive. 

We note that, for the universal conformal mass parameter $M$ in \cref{eq:confL}, the $\phi$ self-interactions can receive effects from both $\chi$s and the SM particles exchanged in the loop. Since in most of the parameter space in \cref{fig:cointeracting}, $\chi$ is heavier than all the SM species, the other contributions are typically suppressed. In particular, for the SM fermions, we expect a $(m_{f}/m_{\chi})^4$ suppression. The only exception can be the SM contribution from the top quark, which can add to the total self-interaction rate of $\phi$ for $m_\chi\sim m_t$.

The loop-induced interaction in \cref{eq:confLapprox} can then be seen as a non-standard kinetic term for the scalar field
\begin{equation}
K(X) = X+a X^2+\ldots\ ,
\label{eq:nonstandkin}
\end{equation}
where $X = -(\partial\phi)^2/2$ and $a \sim m_\chi^4/M^8$. In the non-relativistic limit, we can further write
\begin{equation}
\phi = \frac{1}{\sqrt{2\,m_\phi}}\left(\psi\,e^{-im_\phi t} + \textrm{c.c}\right),
\label{eq:phinonrelativistic}
\end{equation}
where we have denoted the non-relativistic wave function by $\psi$. By substituting \cref{eq:phinonrelativistic} into \cref{eq:confLapprox}, we obtain
\begin{equation}
\mathcal{L} \supset -\alpha(\Lambda)\,\frac{m_\chi^4\,m_\phi^2}{M^8}\,|\psi|^4.
\end{equation}
Notably, in this regime, the impact of the non-standard kinetic term in \cref{eq:nonstandkin} is equivalent to the contribution that one would obtain from the $\lambda\phi^4$ operator with $\lambda>0$ characteristic for repulsive $\phi$ self interactions~\cite{Brax:2019fzb}. Averaging over the oscillatory contributions of the scalar field, the latter interaction would lead to the coupling driven by the $(\lambda/m_\phi^2)\,|\psi|^4$ term. In the model under study, the equivalent tiny coupling constant induced by the non-standard kinetic term can then be written as
\begin{align}
\label{eq:lambdaestimate}
\lambda & = \alpha\frac{m_\chi^4\,m_\phi^4}{M^8}\\
& \sim 10^{-100}\left(\frac{300~\gev}{M}\right)^8\left(\frac{m_\chi}{300~\gev}\right)^4\left(\frac{m_\phi}{3\times 10^{-14}~\textrm{eV}}\right)^4.\nonumber
\end{align}
For the typical values of the model parameters, as indicated in \cref{eq:lambdaestimate}, the obtained $\phi$ self-interaction cross section is then of the order of~\cite{Bento:2000ah,McDonald:2001vt}
\begin{widetext}
\begin{equation}
\frac{\sigma}{m_\phi} = \frac{9}{8\pi}\,\frac{\lambda^2}{m_\phi^3} \sim 10^{-137}\,\left(\frac{300~\gev}{M}\right)^{16}\left(\frac{m_\chi}{300~\gev}\right)^8\left(\frac{m_\phi}{3\times 10^{-14}~\textrm{eV}}\right)^5\ [\cm^2/\g].
\end{equation}
\end{widetext}
As can be seen, it remains much suppressed and can support the creation of the solitons only at very small scales with their typical size given by~\cite{Goodman:2000tg,Arbey:2003sj}
\begin{equation}
r = \sqrt{\frac{3\lambda}{2}}\,\frac{M_{\textrm{Pl}}}{m_\phi^2}\sim 1~\cm\,\left(\frac{300~\gev}{M}\right)^4\left(\frac{m_\chi}{300~\gev}\right)^2\ .
\end{equation}
Interestingly, the characteristic size of the solitons obtained due to $\phi$ self-interactions does not depend on the mass of ultralight scalar $m_\phi$. This is due to the cancellation between the mass dependence of the coupling constant $\lambda$ and such a dependence in the expression for the soliton size $r$ above. In this case, the lighter $\phi$s have smaller loop-induced couplings due to their diminishing characteristic collision energy. This suppresses the creation of larger solitons, which could otherwise be expected for decreasing $m_\phi$. 

As a result, in this effective theory, the solitons with astrophysically relevant sizes of order kpc can be obtained only for a very low conformal mass scale, $M\sim \textrm{tens of $\ev$}$. This lies much below the LHC and LEP bounds; see, however, the discussion in \cref{sec:model} about the validity of these bounds in the limit of low $M$. Instead, for $M\sim \textrm{a few }\times 100~\gev$, the expected impact of the loop-induced self-interaction cross section is negligible. 

Last but not least, we note that even for such a large value of $M$, solitons of a much larger size, can be generated in the dense regions of galaxies, due to the interplay between the gravitational attraction and the quantum pressure, similar to the fuzzy DM scenario. For the masses of  interest, however, and for  velocities of order $(10-100)~\km/\textrm{s}$, we obtain the typical  soliton size $r_{\textrm{quantum}}\sim (10^{-6}-10^{-5})~\textrm{pc}$, which is also of no astrophysical relevance.

\bibliography{main}

\end{document}